\documentclass[pdflatex,sn-mathphys-num]{sn-jnl}

\usepackage{graphicx}
\usepackage{multirow}
\usepackage{amsmath,amssymb,amsfonts}
\usepackage{amsthm}
\usepackage{mathrsfs}
\usepackage[title]{appendix}
\usepackage{xcolor}
\usepackage{textcomp}
\usepackage{manyfoot}
\usepackage{booktabs}
\usepackage{algorithm}
\usepackage{algorithmicx}
\usepackage{algpseudocode}
\usepackage{listings}
\usepackage{tikz}
\usetikzlibrary{arrows.meta,positioning,fit}

\newcommand{\openone}{\mathbb{I}} 

\theoremstyle{thmstyleone}

\theoremstyle{thmstyletwo}

\theoremstyle{thmstylethree}

\begin{document}

\title[Distinguishing synthetic unravelings on quantum computers]{Distinguishing synthetic unravelings on quantum computers}


\author*[1]{\fnm{Eloy} \sur{Pi\~nol}}\email{eloy.p.j@hotmail.com}

\author[2]{\fnm{Piotr} \sur{Sierant}}

\author[3]{\fnm{Dustin} \sur{Keys}}

\author[1]{\fnm{Romain} \sur{Veyron}}

\author[4]{\fnm{Miguel Angel} \sur{Garc\'{\i}a-March}}

\author[5]{\fnm{Tanner} \sur{Reese}}

\author[1,6]{\fnm{Morgan W.} \sur{Mitchell}}

\author[5]{\fnm{Jan} \sur{Wehr}}

\author[1]{\fnm{Maciej} \sur{Lewenstein}}

\affil[1]{\orgname{ICFO -- Institut de Ci\`encies Fot\`oniques, The Barcelona Institute of Science and Technology}, \orgaddress{\city{08860 Castelldefels (Barcelona)}, \country{Spain}}}

\affil[2]{\orgname{Barcelona Supercomputing Center}, \orgaddress{\city{Barcelona 08034}, \country{Spain}}}

\affil[3]{\orgdiv{Independent Researcher}, \orgaddress{\city{Tucson, AZ}, \country{USA}}}

\affil[4]{\orgname{Instituto Universitario de Matem\'atica Pura y Aplicada, Universitat Polit\`ecnica de Val\`encia}, \orgaddress{\street{Camino de Vera, s/n}, \city{46022 Valencia}, \country{Spain}}}

\affil[5]{\orgdiv{Department of Mathematics}, \orgname{The University of Arizona}, \orgaddress{\city{Tucson, AZ 85721-0089}, \country{USA}}}

\affil[6]{\orgname{ICREA -- Instituci\'o Catalana de Recerca i Estudis Avan\c{c}ats}, \orgaddress{\city{08010 Barcelona}, \country{Spain}}}

\abstract{
Distinct monitoring or intervention schemes can produce different conditioned stochastic quantum trajectories while sharing the same unconditional (ensemble-averaged) dynamics. This is the essence of unravelings of a given Gorini--Kossakowski--Sudarshan--Lindblad (GKSL) master equation: any trajectory-ensemble average of a function that is linear in the conditional state is completely determined by the unconditional density matrix, whereas applying a nonlinear function before averaging can yield unraveling-dependent results and thereby reveal measurement backaction beyond the average evolution. A paradigmatic example is resonance fluorescence, where direct photodetection (jump/Poisson) and homodyne or heterodyne detection (diffusive/Wiener) define inequivalent unravelings of the same GKSL dynamics. In earlier work, we showed that nonlinear trajectory averages can distinguish such unravelings, but observing the effect in that optical setting requires demanding experimental precision. Here we translate the same idea to a digital, highly controlled setting by introducing \emph{synthetic unravelings} implemented as quantum circuits acting on one and two qubits. We design two unravelings---a projective measurement unraveling and a random-unitary ``kick'' unraveling---that share the same ensemble-averaged evolution while yielding different nonlinear conditional-state statistics. We implement the protocols on superconducting-qubit hardware provided by IBM Quantum, combining circuit design, readout-error mitigation, and classical post-processing to access trajectory-level information. We show that the variance across trajectories and the ensemble-averaged von Neumann entropy distinguish the unravelings in both theory and experiment, while the unconditional state and the ensemble-averaged expectation values that are linear in the state remain identical. Our results provide an accessible demonstration that quantum trajectories encode information about measurement backaction beyond what is fixed by the unconditional dynamics.
}


\maketitle

\section{Introduction}\label{sec:intro}

Open quantum systems are commonly described through a reduced density operator \(\rho_t\), obtained after tracing out environmental degrees of freedom. Under Markovian assumptions, the evolution of \(\rho_t\) is generated by the Gorini--Kossakowski--Sudarshan--Lindblad (GKSL) master equation~\cite{Lindblad1976,Gorini1976}:
\begin{equation}
  \dot{\rho}_t = -i[H_S,\rho_t]
  + \sum_k \gamma_k \left( L_k \rho_t L_k^\dagger - \tfrac{1}{2}\{L_k^\dagger L_k,\rho_t\} \right),
  \label{eq:Lindblad}
\end{equation}
where \(H_S\) is the system Hamiltonian, \(L_k\) are Lindblad operators, and \(\gamma_k\) are the associated rates.

If the environment is monitored, the system state conditioned on the measurement record \(r\) follows a stochastic evolution \(\rho_t^{(r)}\), often referred to as a quantum trajectory~\cite{Carmichael1993, Dalibard1992, WisemanMilburn,WisemanDiosi}. Different monitoring schemes define different \emph{unravelings} of the same GKSL dynamics, in the sense that they generate different ensembles of conditional states \(\{\rho_t^{(r)}\}\) whose average reproduces the same unconditional evolution,
\begin{equation}
  \mathbb{E}_r\!\left[\rho_t^{(r)}\right] \equiv \overline{\rho_t^{(r)}} = \rho_t .
\end{equation}

A key point is that the master equation fixes \(\rho_t\), but it does not fix the conditional dynamics. As a consequence, any quantity that depends only on \(\rho_t\)---for example, any expectation value that is linear in the state such as \(\langle O\rangle=\mathrm{Tr}(O\rho_t)\)---is insensitive to the unraveling. Unraveling dependence appears when one probes the \emph{distribution} of conditional states through trajectory-ensemble averages of nonlinear functions of the conditional state (hereafter, for brevity, \emph{nonlinear trajectory averages}), such as variances across trajectories or trajectory-averaged entropies. 

In a previous work~\cite{Pinol2024unravelings}, we showed that the trajectory variance of conditional expectation values distinguishes inequivalent unravelings in the resonance fluorescence problem~\cite{Mollow69}, also characterizing its robustness against detector inefficiencies and thermal noise. 
However, an experimental verification of these findings is highly nontrivial as it requires nearly perfect detection efficiencies to faithfully access the measurement record and thus resolve the intrinsic trajectory-to-trajectory fluctuations; otherwise, imperfect detection and additional noise sources (e.g., thermal photons) wash out the conditioning and strongly suppress the trajectory variance.

\begin{figure}[t!]
\centering
\includegraphics[width=\textwidth]{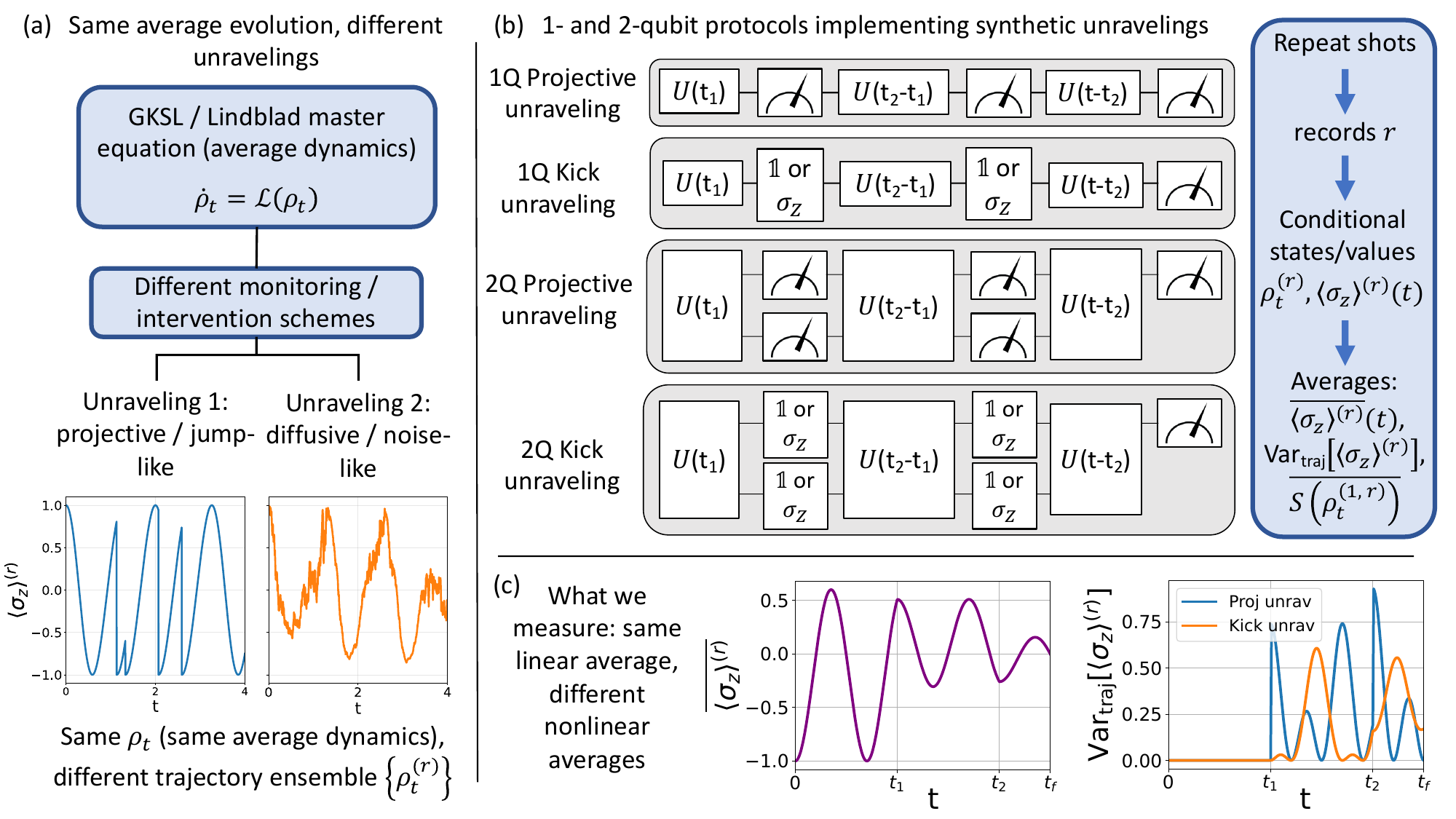}
\caption{\textbf{Conceptual overview of this work.}
\textbf{(a)} \textbf{Theoretical framework:} A GKSL master equation fixes the average evolution \(\dot{\rho}_t=\mathcal{L}(\rho_t)\) but leaves the conditional dynamics undefined. Different monitoring/intervention schemes correspond to inequivalent unravelings, generating distinct trajectory ensembles \(\{\rho_t^{(r)}\}\).
\textbf{(b)} \textbf{Quantum-circuit protocol:} We engineer \emph{synthetic unravelings} with discrete quantum circuits on one and two qubits. The unitary evolution is interrupted by interventions---either projective measurements or random-unitary ``kicks'' (applying \(\openone\) or \(\sigma_z\) with 50\% probability)---that generate distinct trajectory ensembles while preserving the same average state, \(\rho_t\). The flowchart (right) summarizes the workflow: repeating shots to collect records \(r\), reconstructing conditional states/values, and computing both linear and nonlinear statistics.
\textbf{(c)} \textbf{Distinguishing unravelings:} Unraveling dependence appears only in nonlinear trajectory averages. While the linear average \(\overline{\langle\sigma_z\rangle^{(r)}}(t)\) is identical for both schemes, the trajectory variance \(\mathrm{Var}_\mathrm{traj}[\langle\sigma_z\rangle^{(r)}]\) depends on the unraveling, revealing the intervention- (or monitoring-) induced conditioning at the trajectory level (data shown for the 1Q protocols).}
\label{fig:conceptual}
\end{figure}

Here, we bypass these experimental limitations by translating the concept to a digital, highly controlled setting: \emph{synthetic unravelings} implemented as discrete quantum-circuit protocols on a quantum processor. Instead of realizing resonance fluorescence, we implement a protocol for one and two qubits that retains the same logical structure: a fixed average evolution together with two inequivalent unravelings. We implement the protocol on IBM Quantum superconducting hardware, combining circuit design, readout-error mitigation, and classical post-processing to access trajectory-level information (Fig.~\ref{fig:conceptual}). Combining repeated circuit executions, we can distinguish the unravelings via the trajectory-averaged variance and the trajectory-averaged von Neumann entropy, providing a clear demonstration of measurement backaction effects that are invisible to linear average dynamics.

\section{Master equations, unravelings and nonlinear trajectory averages}\label{sec:unravelings}

\subsection{Unravelings of a GKSL master equation}

An open quantum system interacts with its environment, and together they undergo a joint unitary evolution. The interaction generally leads to entanglement~\cite{Horodecki09} between the system and environmental degrees of freedom, which the continuous monitoring of the environment can reduce or even destroy~\cite{Nha04, Carvalho07, Barchielli13entanglement, Guevara14, Cao2019SciPost,  Barchielli24}. Since quantum measurements yield intrinsically random outcomes, this monitoring induces a stochastic dynamics on the system state. Crucially, distinct monitoring schemes (corresponding to measurements in different bases) yield different stochastic trajectories, thereby defining different \emph{unravelings} of the same reduced dynamics.

An unraveling can be formulated either as a stochastic Schr\"odinger equation for pure conditional states \(|\psi_t^{(r)}\rangle\) or as a stochastic master equation for mixed conditional states \(\rho_t^{(r)}\)~\cite{Carmichael1993,WisemanMilburn,WisemanDiosi,JacobsSteck}. In quantum optics, these are commonly categorized into two archetypes:
\begin{itemize}
  \item \emph{Jump (Poisson) unravelings}, associated with discrete detection events such as photon counts~\cite{Minev2019};
  \item \emph{Diffusive (Wiener) unravelings}, associated with continuous measurement currents such as homodyne or heterodyne detection~\cite{BarchielliGregoratti}.
\end{itemize}
These categories are not mutually exclusive; general unravelings can involve combinations of both Poissonian jumps and Wiener noise, depending on the detection scheme (e.g., inefficient photodetection or general-dyne detection). In all cases, averaging over the ensemble of records recovers the unconditional GKSL evolution given by Eq.~\eqref{eq:Lindblad}.

Importantly, quantum trajectories play a dual role. Physically, they represent the conditioned state of a system under a specific monitoring scheme. Computationally, when corresponding to pure conditional states $|\psi_t^{(r)}\rangle$, they serve as a powerful numerical tool known as the Monte Carlo Wave Function method~\cite{Dalibard1992, PlenioKnight1998, BreuerPetruccione}. In this computational context, one often chooses a specific unraveling -- typically based on computational efficiency~\cite{Vovk22} --- regardless of whether it corresponds to the actual physical interaction or detection setup.

\subsection{Nonlinear functions on trajectory ensembles}

At a fixed time \(t\), an unraveling defines a trajectory ensemble \(\mathcal{E}_t=\{\rho_t^{(r)},p(r)\}\), where \(p(r)\) denotes the probability (or probability density) of the measurement record \(r\). While all unravelings compatible with a given master equation share the same unconditional state \(\rho_t=\mathbb{E}_r[\rho_t^{(r)}]\), they can generate different \emph{distributions} of conditional states. This difference can only be detected when one probes quantities that depend on more than the average state.

Indeed, for any function \(F(\rho)\) that is \emph{linear} in the state,
\begin{equation}
\mathbb{E}_r\!\left[F(\rho_t^{(r)})\right]
=F\!\left(\mathbb{E}_r[\rho_t^{(r)}]\right)
=F(\rho_t),
\label{eq:linear-commutes}
\end{equation}
where the last equality shows that the trajectory average of the values \(F(\rho_t^{(r)})\) coincides with evaluating \(F\) on the unconditional state \(\rho_t\).
Therefore, if two unravelings share the same \(\rho_t\), they necessarily yield identical values for \(\mathbb{E}_r[F(\rho_t^{(r)})]\), and such quantities cannot distinguish them.
In contrast, \emph{nonlinear} functions generally do not satisfy Eq.~\eqref{eq:linear-commutes}: the trajectory-ensemble average of a nonlinear function of the conditional state, \(\mathbb{E}_r[F(\rho_t^{(r)})]\), is no longer fixed by \(\rho_t\) and can therefore depend on the unraveling:
\begin{equation}
F(\rho_t) \neq \mathbb{E}_r\!\left[F(\rho_t^{(r)})\right].
\end{equation}
That is, applying \(F\) to the averaged state differs from applying \(F\) to each \(\rho_t^{(r)}\) and then taking the average over trajectories. This noncommutativity is precisely what allows nonlinear trajectory-ensemble averages to reveal unraveling dependence, and it has deep consequences.

A simple example is the variance across trajectories of the conditional expectation value of an observable \(O\):
\begin{equation}
  \mathrm{Var}_\mathrm{traj}\big[\langle O \rangle\big]
  = \mathbb{E}_r \left[ \left( \langle O \rangle^{(r)}\right)^2 \right]
    - \left( \mathbb{E}_r \left[ \langle O \rangle^{(r)} \right] \right)^2,
  \label{eq:traj-variance}
\end{equation}
where \(\langle O \rangle^{(r)}=\mathrm{Tr}\!\left(O\rho_t^{(r)}\right)\) is the conditional expectation value at time \(t\) along record \(r\).
The second term is fixed by the unconditional dynamics, while the first one captures trajectory fluctuations and can depend on the unraveling choice.

Trajectory ensembles also allow one to define entropy-based probes~\cite{VovkPichler2024,KeysWehr24NonlinearFunctionals}, which are nonlinear functions of the state. By concavity of the von Neumann entropy \(S(\rho)\equiv-\mathrm{Tr}(\rho\log_2\rho)\)~\cite{NielsenChuang2010,LiebRuskai1973}, we find
\begin{equation}
    S(\rho_t) \;\ge\; \mathbb{E}_r\!\left[S\!\left(\rho_t^{(r)}\right)\right].
\end{equation}

The gap between these two quantities reflects information present at the trajectory level that is not fixed by \(\rho_t\) alone. Other nonlinear functions and higher-order correlation measures have been used to characterize trajectory ensembles~\cite{Nha04,VovkPichler2024,Verstraelen2023}. Recent work has also discussed limits on protocols that aim to distinguish unravelings associated with the same master equation~\cite{GaonaReyes2025}, highlighting that operational distinguishability relies on access to the labels that identify each trajectory.

\section{Resonance fluorescence and unravelings of a two-level atom dynamics}\label{sec:RF}

Resonance fluorescence of a driven two-level atom coupled to the electromagnetic field provides a standard setting in which a single master equation admits multiple inequivalent unravelings~\cite{WisemanMilburn}. The atom is modeled as a two-level system with ground state \(|g\rangle\) and excited state \(|e\rangle\). In the interaction picture with respect to the drive, and within the rotating-wave approximation, the Hamiltonian reads
\begin{equation}
H_\mathrm{RF} = \tfrac{\Omega}{2}\,\sigma_x - \tfrac{\Delta}{2}\,\sigma_z,
\label{eq:H_RF}
\end{equation}
where \(\Omega\) is the Rabi frequency, \(\Delta\) is the detuning, \(\sigma_{x,z}\) are Pauli matrices, and \(\hbar = 1\). Radiative decay at rate \(\gamma\) is described by the Lindblad operator \(L = \sqrt{\gamma}\,\sigma_-\).
The reduced atomic state \(\rho_t\) obeys
\begin{equation}
\dot{\rho}_t = -i\big[H_\mathrm{RF},\rho_t\big]
+ \gamma\left(\sigma_- \rho_t \sigma_+ - \tfrac{1}{2}\{\sigma_+\sigma_-,\rho_t\}\right),
\label{eq:RF_ME}
\end{equation}
where $\sigma_\pm = \tfrac{1}{2}(\sigma_x \pm i \sigma_y)$.
The master equation~\eqref{eq:RF_ME} describes the unconditional dynamics of the atom (i.e., when the output field is not monitored), which is equivalent to the average over the trajectories associated with any unraveling.

Specifically, distinct schemes for monitoring the emitted field define different unravelings of the same reduced atomic dynamics~\cite{Carmichael1993,WisemanMilburn,WisemanDiosi,JacobsSteck,PlenioKnight1998,Dalibard1992}. Two standard examples are:
\begin{itemize}
\item \emph{Direct photodetection}: the field is monitored by an ideal photon counter. The record is a point process and the conditional state undergoes jumps. The corresponding stochastic Schr\"odinger equation is driven by a Poisson process, with non-Hermitian evolution between clicks and updates proportional to \(\sigma_-\) at detection times.
\item \emph{Homodyne detection}: the field is mixed with a strong local oscillator and measured with a balanced homodyne receiver. The record is a continuous photocurrent and the conditional state follows a diffusive stochastic Schr\"odinger equation driven by a Wiener process.
\end{itemize}
Explicit forms of the jumps and diffusive stochastic equations can be found in Refs.~\cite{Carmichael1993,WisemanMilburn,WisemanDiosi,JacobsSteck,Dalibard1992,PlenioKnight1998}. In both cases, the unconditional state satisfies \(\rho_t=\mathbb{E}_r[\rho_t^{(r)}]\) and obeys Eq.~\eqref{eq:RF_ME}. As a result, any trajectory-ensemble average of a linear function of the conditional state has the same time dependence for the two monitoring schemes, including expectation values \(\mathbb{E}_r \big[ \langle O\rangle^{(r)} \big] = \langle O\rangle\).

To distinguish them, one must turn to nonlinear trajectory averages. A convenient probe is the variance across trajectories of the atomic inversion expectation value~\cite{Pinol2024unravelings}. Applying the general definition~\eqref{eq:traj-variance} to \(\langle \sigma_z \rangle\), we obtain:
\begin{equation}
\mathrm{Var}_\mathrm{traj}\big[\langle \sigma_z \rangle\big]
= \mathbb{E}_r\left[\left( \langle \sigma_z \rangle^{(r)}\right)^2 \right]
- \left(\mathbb{E}_r\left[ \langle \sigma_z \rangle^{(r)} \right]\right)^2,
\label{eq:RF_traj_var}
\end{equation}
where \(\langle \sigma_z \rangle^{(r)}=\mathrm{Tr}(\sigma_z\rho_t^{(r)})\). The variance \(\mathrm{Var}_\mathrm{traj}\big[\langle \sigma_z \rangle\big]\) depends on the full distribution of conditional values and can therefore distinguish unravelings even when the ensemble-averaged state is the same. However, turning this theoretical prediction into a laboratory test is experimentally demanding. As analyzed in Ref.~\cite{Pinol2024unravelings}, the trajectory variance is particularly sensitive to imperfect detection and thermal noise, which tend to wash out the information carried by the record and reduce the contrast between trajectory ensembles.

These constraints motivate the following considerations of protocols that retain the key ingredients---a fixed average evolution, multiple inequivalent unravelings, and nonlinear trajectory averages that distinguish them---but replace continuous monitoring by operations available in gate-based platforms. This is the purpose of the protocols introduced below: unitary continuous evolution is interrupted at fixed times by interventions that implement the same dephasing channel at the level of the average state, while generating different trajectory ensembles. We implement single- and two-qubit versions of this construction on quantum hardware using only unitary gates, projective measurements, and classical post-processing, and we extract the trajectory-level statistics from repeated runs to distinguish the unravelings.

\section{Synthetic unravelings: single- and two-qubit protocols}\label{sec:RF-to-synthetic}

We introduce single- and two-qubit protocols that reproduce the structure of inequivalent unravelings in discrete time, in the spirit of discrete-time trajectory models~\cite{Brun2002}, that can be run on quantum hardware (Fig.~\ref{fig:conceptual}b).

Both protocols share the same structure:
\begin{itemize}
\item unitary evolution segments generated by a fixed Hamiltonian, separated by two intervention times \(t_1\) and \(t_2\);
\item at each intervention, we apply either a projective measurement or a random unitary operation (applying identity or a phase-flip $\sigma_z$ with equal probability). Both options induce the same dephasing of the average state in the computational basis;
\item a final projective measurement of \(\sigma_z\) on a designated qubit, used to reconstruct the trajectory ensemble and compute nonlinear quantities.
\end{itemize}

The average state and all quantities obtained from it are the same in the two unravelings. Differences arise in trajectory averages of nonlinear functions of the conditional states \(\rho_t^{(r)}\).

Because the interventions occur at fixed times, the trajectory ensemble is discrete and its size depends on how many interventions have taken place by the final time \(t\). Denoting by \(m(t)\in\{0,1,2\}\) the number of interventions completed by time \(t\), the single-qubit protocol yields \(2^{m(t)}\) possible trajectories in both unravelings (measurement outcomes \(b\in\{0,1\}\) or kicks \(I/Z\)). Likewise, the two-qubit protocol yields \(4^{m(t)}\) possible trajectories in both unravelings (joint outcomes \(b_1b_2\in\{00,01,10,11\}\) or kick patterns \(K_{ab}\)). In particular, for \(t_1<t<t_2\) there are \(2\) (single qubit) and \(4\) (two qubits) trajectories, while for \(t>t_2\) there are \(4\) and \(16\), respectively.

\subsection{Single-qubit protocol}\label{subsec:1Q_synthetic}

The system Hilbert space is spanned by \(\{\lvert 0\rangle,\lvert 1\rangle\}\). We use the same Hamiltonian Eq.~\eqref{eq:H_RF} as in the resonance fluorescence case for the continuous evolution of the qubit between interventions. The circuit implements the unitary evolution generated by \(e^{-iH_\mathrm{RF}t}\). We initialize the system in \(\lvert\psi_0\rangle = \lvert 0\rangle\) and adopt the convention \(\sigma_z \lvert 0\rangle = -\lvert 0\rangle\), resulting in that \(\overline{\langle \sigma_z \rangle^{(r)}}(0) = -1\).

At times \(t_1\) and \(t_2\) we implement one of two realizations of the same average dynamics in the computational basis (Fig.~\ref{fig:conceptual}b):
\begin{itemize}
\item \emph{Projective unraveling:} at each intervention we perform a projective measurement in the computational basis \(\{\lvert 0\rangle,\lvert 1\rangle\}\). This results in a fully dephasing channel to the average state.
\item \emph{Kick unraveling:} at each intervention we randomly apply either the identity or a phase-flip (\(\sigma_z\)) operation, each with probability \(1/2\). This also produces a fully dephasing channel in the computational basis.
\end{itemize}

\begin{figure}[t!]
 \centering
 \includegraphics[width=\columnwidth]{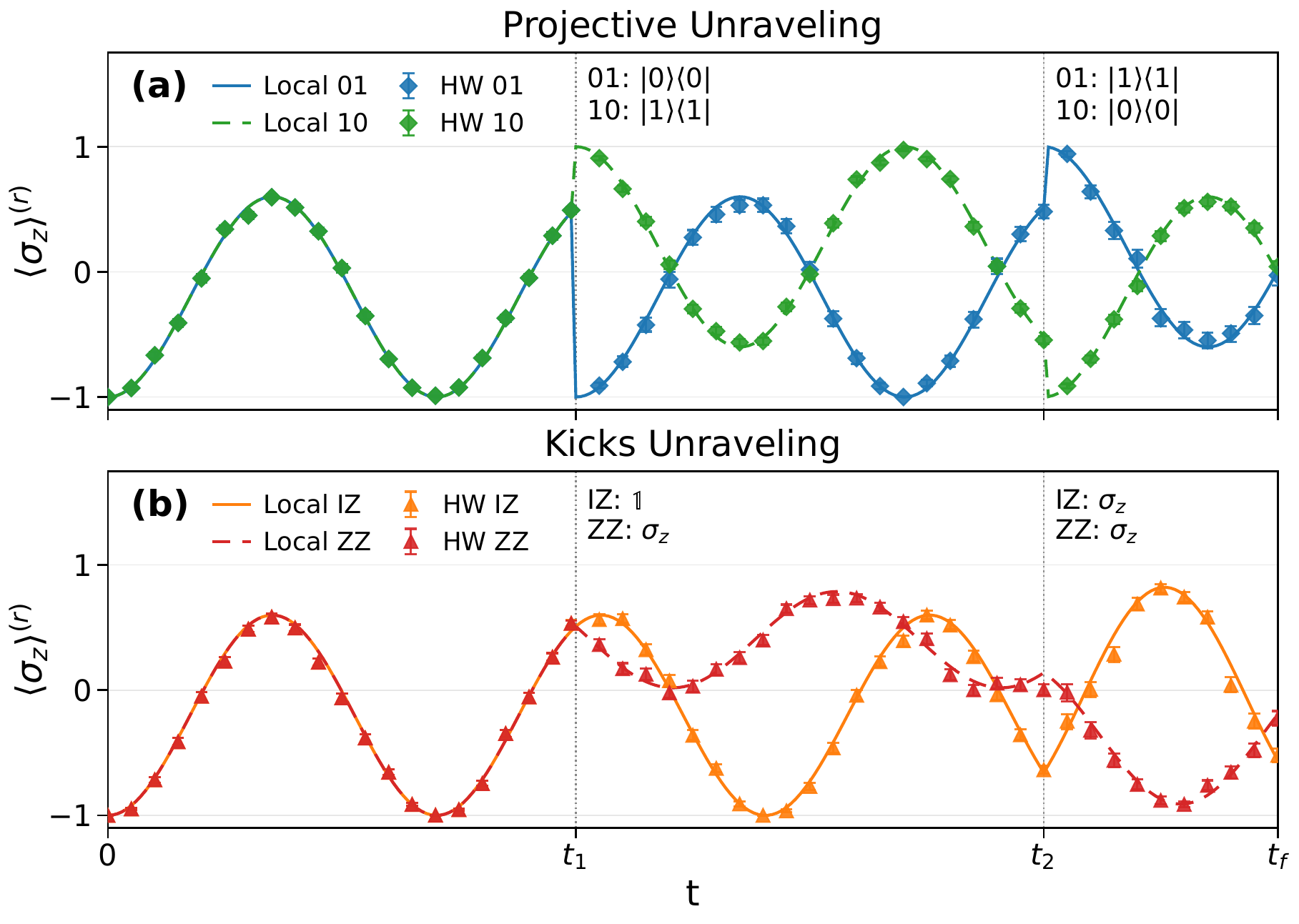}
\caption{Single-qubit trajectories on quantum hardware.
\textbf{(a)} Projective unraveling: \(\langle \sigma_z \rangle^{(r)}\) changes discontinuously at \(t_1\) and \(t_2\) due to the state update conditioned on the measurement outcome. The annotations next to the vertical intervention lines indicate the projector associated with that trajectory branch. Legend labels denote the sequence of outcomes (e.g., ``01'' corresponds to outcome \(0\) at \(t_1\) and \(1\) at \(t_2\)).
\textbf{(b)} Kick unraveling: the state evolves unitarily, with updates at the intervention times. The annotations indicate the applied gate (\(\openone\) or \(\sigma_z\)). Legend labels denote the sequence of kicks (e.g., ``IZ'' corresponds to \(\openone\) at \(t_1\) and \(\sigma_z\) at \(t_2\)).
Lines show theory and markers show experimental data. Error bars are obtained via bootstrap resampling of the shot counts at each time point.}
\label{fig:trajectories_comparison}
\end{figure}

Figure~\ref{fig:trajectories_comparison} illustrates the distinct trajectory dynamics of each unraveling. In the projective case, interventions cause instantaneous jumps in \(\langle \sigma_z \rangle^{(r)}\) due to state collapse. In contrast, the kick unraveling preserves the continuity of \(\langle \sigma_z \rangle^{(r)}\); the applied unitaries do not discontinuously shift the expectation value but rather can alter the state's phase, redirecting the subsequent evolution. Despite these drastic differences at the trajectory level, the average state evolution remains identical for both schemes.

\subsection{Two-qubit protocol}\label{subsec:2Q_synthetic}

The two-qubit protocol extends the construction to a pair of interacting qubits. The Hilbert space is spanned by \(\{\lvert b_1 b_2\rangle : b_j \in \{0,1\}\}\). We take
\begin{equation}
H = H_\mathrm{RF} \otimes \openone + \openone \otimes H_\mathrm{RF} + J\,\sigma_z \otimes \sigma_z,
\label{eq:H-two-qubits}
\end{equation}
where \(J\) is the coupling constant and \(\openone\) denotes the \(2\times 2\) identity. This Hamiltonian generates the continuous unitary evolution of the joint state. The initial state is \(\lvert\psi_0\rangle = \lvert 0 0\rangle\). At times \(t_1\) and \(t_2\) we again implement two realizations of the same dephasing dynamics in the computational basis (Fig.~\ref{fig:conceptual}b):

\begin{itemize}
\item \emph{Projective unraveling:} we perform a joint projective measurement of both qubits in the computational basis \(\{\lvert b_1 b_2\rangle\}\), effectively removing all coherence between basis states.
\item \emph{Kick unraveling:} we independently apply either the identity or a phase-flip (\(\sigma_z\)) gate to each qubit with probability \(1/2\). At each intervention we apply one of the four commuting unitaries
\begin{equation}
K_{ab} = \sigma_z^{a} \otimes \sigma_z^{b}, \qquad a,b \in \{0,1\},
\end{equation}
each with probability \(1/4\).
\end{itemize}
In both versions, the final readout used to construct \(\overline{\langle \sigma_z \rangle^{(r)}}(t)\) and \(\mathrm{Var}_\mathrm{traj}[\langle \sigma_z \rangle^{(r)}](t)\) is a projective measurement of the local Pauli operator $\sigma_z^{(1)} = \sigma_z \otimes \openone$ on the first qubit only.

\subsection{Discrete Kraus maps and shared dephasing channels}

Both the single- and two-qubit protocols are constructed so that, at each intervention time, the projective and kick versions implement the same completely positive trace-preserving (CPTP) dephasing map on the average state. In the projective unraveling, the measurement outcomes (and hence the corresponding trajectory branches) occur with probabilities given by the Born rule, \(p_b=\mathrm{Tr}(P_b\,\rho\,P_b)=\mathrm{Tr}(P_b\rho)\), and the conditional post-measurement state is \(\rho^{(b)}=P_b\rho P_b/p_b\).

For the single-qubit case, the two maps are
\begin{align}
\mathcal{M}^{(1)}_\mathrm{proj}(\rho)
&= \sum_{b=0,1} P_b\,\rho\,P_b, &
P_b &= \lvert b\rangle\langle b\rvert, \\
\mathcal{M}^{(1)}_\mathrm{kick}(\rho)
&= \tfrac{1}{2}\sum_{a=0,1} \sigma_z^a\,\rho\,\sigma_z^a .
\end{align}
Writing \(\rho\) in the basis \(\{\lvert 0\rangle,\lvert 1\rangle\}\), one finds
\begin{equation}
\mathcal{M}^{(1)}_\mathrm{proj}(\rho) = \mathcal{M}^{(1)}_\mathrm{kick}(\rho) = \begin{pmatrix}
\rho_{00} & 0 \\
0 & \rho_{11}
\end{pmatrix}.
\end{equation}
Thus both constructions leave populations unchanged and set coherences to zero.

Similarly, for the two-qubit protocol the projective and kick maps are
\begin{align}
\mathcal{M}^{(2)}_\mathrm{proj}(\rho)
&= \sum_{b_1,b_2\in\{0,1\}} P_{b_1 b_2}\,\rho\,P_{b_1 b_2}, &
P_{b_1 b_2} &= \lvert b_1 b_2\rangle\!\langle b_1 b_2\rvert, \\
\mathcal{M}^{(2)}_\mathrm{kick}(\rho)
&= \tfrac{1}{4}\sum_{a,b\in\{0,1\}} K_{ab}\,\rho\,K_{ab}.
\end{align}
Again, \(\mathcal{M}^{(2)}_\mathrm{proj}(\rho) = \mathcal{M}^{(2)}_\mathrm{kick}(\rho)\), so both unravelings implement the same two-qubit dephasing channel.
Though we only consider CPTP maps for one- and two-qubits here,
for a general CPTP $ \mathcal{M} $, the existence of an unraveling, $ \{K_i\} $, such that $ \mathcal{M}(\rho) = \sum_i K_i \rho K_i^\dagger $
is ensured by the Kraus representation theorem, in the finite dimensional case,
and the Stinespring representation theorem, in infinite dimensions~\cite{Hayashi15QuantumInfo}.
Additionally, the Stinespring representation theorem can be used to determine the conditions
under which two unravelings produce the same CPTP map.

\section{Implementation and data analysis}\label{sec:implementation-analysis}

We implement the single- and two-qubit protocols as families of quantum circuits executed on superconducting-qubit devices provided by IBM Quantum~\cite{IBMQuantum}. For each final time \(t\), the circuit is built from unitary segments separated by the intervention times \(t_1\) and \(t_2\). All compilation and job submission are performed using Qiskit~\cite{Qiskit} together with IBM Quantum Runtime (Sampler).

\paragraph{Circuit construction and interventions.}
For each \(t\), we transpile the target unitary evolution generated by \(H_\mathrm{RF}\) (single qubit) or \(H\) (two qubits) into the native gate set of the chosen backend. At the intervention slots we insert either mid-circuit measurements (projective unraveling) or \(\sigma_z\) unitary kicks (kick unraveling). In the kick unraveling we execute one circuit per kick pattern and combine the shot counts across patterns with their corresponding weights when estimating trajectory-level quantities. At the final time \(t\) we always measure \(\sigma_z\) on the designated qubit (qubit~1).

\paragraph{Readout mitigation.}
We apply readout error mitigation conditioned on the hardware performance. For the single-qubit experiments, the readout fidelity on the ibm\_torino backend was  highly accurate (single-qubit readout fidelity \(F \equiv [P(0|0)+P(1|1)]/2 \approx 99.5\%\)), so mitigation produced negligible changes in all reported quantities. We therefore report raw measurement data for the single-qubit results to avoid unnecessary post-processing.

For the two-qubit experiments, where the combined readout error is significant, we apply a bounded least-squares unfolding method. We calibrate the measurement device by constructing the assignment matrix $M$ via preparation and measurement of the computational basis states. The true probability vector is then estimated from the observed frequencies by minimizing the squared error norm subject to physical constraints (non-negativity and normalization).

\subsection{Nonlinear conditional-state statistics from trajectory ensembles}\label{subsec:observables}

Our main nonlinear probe is the variance across trajectories of the conditional expectation value of the measured local observable (denoted as $\sigma_z$ for the single-qubit case and $\sigma_z^{(1)}$ for the two-qubit case).
For a compact notation, we denote the generic expectation value as $\langle\sigma_z\rangle_t^{(r)}$.
In the projective unraveling, trajectories are defined by the intermediate measurement outcomes at \(t_1\) and \(t_2\). In the kick unraveling, trajectories are defined by the kick pattern applied at those times.

For each time \(t\), we represent the empirical trajectory ensemble as a set of branches \(r\) with weights \(w_r(t)\) and conditional estimates \(e_r(t)\). We denote by \(e_r(t)\) the finite-shot estimator of the conditional expectation value
\begin{equation}
  \langle\sigma_z\rangle_t^{(r)}=\mathrm{Tr}\!\big(O_{\mathrm{meas}}\,\rho_t^{(r)}\big),
\end{equation}
where \(O_{\mathrm{meas}}\) represents the relevant operator (\(\sigma_z\) or \(\sigma_z^{(1)}\)).
The trajectory-averaged expectation value and trajectory variance are
\begin{equation}
  \mu(t) = \sum_r w_r(t)\, e_r(t),
  \qquad
  \mathrm{Var}_\mathrm{traj}(t) = \sum_r w_r(t)\big(e_r(t) - \mu(t)\big)^2.
  \label{eq:mu-vartraj}
\end{equation}

To determine the conditional estimate \(e_r(t)\), we use the \(N_r\) measurement shots assigned to branch \(r\). Each shot yields a binary outcome \(b\in\{0,1\}\), mapped to eigenvalues \(b=0\mapsto -1\) and \(b=1\mapsto +1\).
For the single-qubit case, \(e_r(t)\) is computed directly from the raw relative frequencies.
For the two-qubit case, the raw counts are processed with the calibration matrix to infer the mitigated probabilities \(p_r(0\,|\,t)\) and \(p_r(1\,|\,t)\), yielding the estimator \(e_r(t) = p_r(1\,|\,t) - p_r(0\,|\,t)\).

By construction, \(\mu(t)\) coincides with the expectation value of the average state, \(\mu(t)=\mathrm{Tr}(O_{\mathrm{meas}}\,\rho_t)\), while \(\mathrm{Var}_\mathrm{traj}(t)\) is unraveling dependent.

\paragraph{Projective unraveling.}
For each shot we record the intermediate outcomes at \(t_1\) and \(t_2\) (when applicable) and the final outcome of qubit~1. We group shots by their intermediate outcomes, treating each group as a branch \(r\). The branch weight \(w_r(t)\) is the corresponding empirical frequency. For each branch we estimate the probabilities for the final measurement of qubit~1 (applying readout mitigation only for the two-qubit protocol) and compute \(e_r(t)\).

\paragraph{Kick unraveling.}
Each branch \(r\) corresponds to a fixed kick pattern, implemented as a separate circuit. For each circuit we estimate the final probabilities for qubit~1 (again, mitigating readout errors only for the two-qubit case) and obtain \(e_r(t)\). The weights \(w_r(t)\) are ideally uniform; in practice we use the executed shot counts per circuit and normalize across patterns.

\paragraph{Error bars via bootstrap.}
All experimental error bars are obtained by bootstrap resampling at each time point \(t\). For single-qubit data, we resample the raw shot counts and compute the estimators directly. For two-qubit data, the resampling is performed on the raw counts prior to the unfolding step to propagate the uncertainty of the mitigation process. Concretely, for each \(t\) we generate synthetic datasets by resampling the observed counts within each branch (projective) or circuit (kick), recompute the estimator, and extract the \(16\)th and \(84\)th percentiles as the central \(68\%\) confidence interval.

\paragraph{Entropies.}
In addition to the trajectory variance, we consider von Neumann entropies associated with qubit~1. At the trajectory level we define
\begin{equation}
  \mathbb{E}_r\!\big[\,S\big(\rho^{(1,r)}_t\big)\big],
  \label{eq:ent1}
\end{equation}
where \(\rho^{(1,r)}_t\) is the reduced state of qubit~1 along trajectory \(r\). 
In the experimental implementation we do not perform full state tomography along each branch. Therefore, the entropy data shown with markers in Sec.~\ref{sec:results} are \emph{semi-experimental}: we use the empirical branch weights \(w_r(t)\) extracted from the measured trajectory records of the projective unraveling, while the branch states \(\rho^{(1,r)}_t\) entering \(S(\cdot)\) are taken from the corresponding ideal (theoretical) evolution of the protocol. 

In contrast, the entropy of the ensemble-averaged reduced state,
\begin{equation}
  S\!\big(\rho^{(1)}_t\big),
  \qquad
  \rho^{(1)}_t = \mathrm{Tr}_2\,\rho_t,
  \label{eq:ent2}
\end{equation}
depends only on the ensemble-averaged evolution and is therefore identical for all unravelings that share the same average dynamics. We note that the entropies evaluated at the trajectory level, Eq.~\eqref{eq:ent1}, quantify the entanglement between qubit~1 and the rest of the system in the pure conditioned state $| \psi_t^{(r)} \rangle$~\cite{Plenio06ent},
whereas \(S\!\big(\rho^{(1)}_t\big)\) characterizes only the lack of purity of the ensemble-averaged reduced state and is not an entanglement measure in the considered setting.

\section{Results: distinguishing unravelings on quantum hardware}\label{sec:results}

\subsection{Single-qubit benchmark on quantum processor}\label{subsec:1Q_results}

The single-qubit protocol serves as a baseline where no inter-qubit correlations exist. Yet, it already illustrates the main message of this work: linear average dynamics can coincide in two different scenarios which differ by the nonlinear averages.

We implement the protocol of Sec.~\ref{subsec:1Q_synthetic}. At the level of the average density operator, both constructions realize the same dephasing channel at the intervention times, and therefore the same average evolution. At the trajectory level, however, the conditional updates differ, and so do unraveling-sensitive nonlinear conditional-state statistics. We have checked in local simulations that the coexistence of identical ensemble averages and distinct trajectory-level signatures is robust for different values of the protocol parameters.

\begin{figure}[t]
 \centering
 \includegraphics[width=\linewidth]{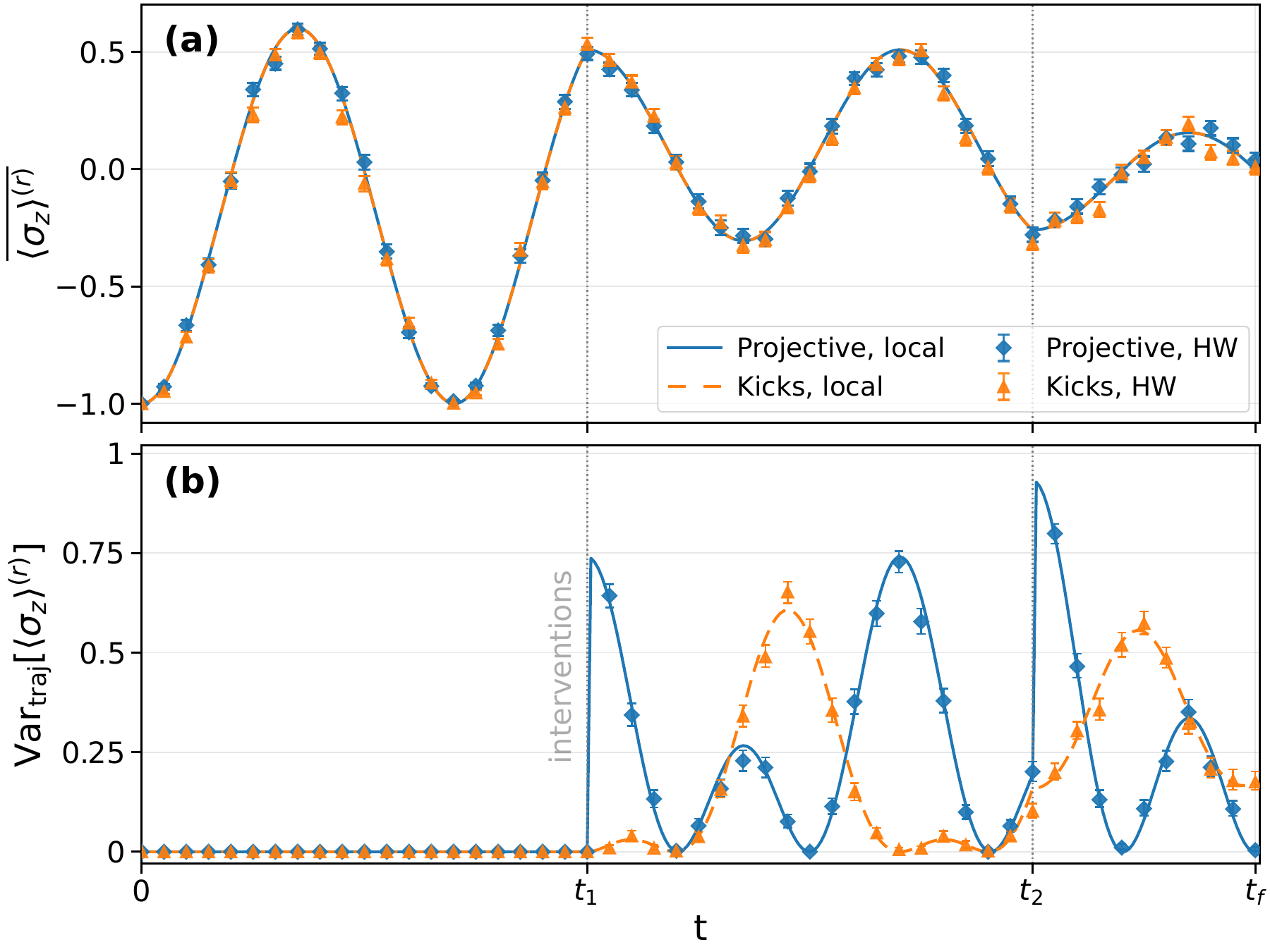}
\caption{Single-qubit protocol.
\textbf{(a)} Trajectory-averaged expectation value \(\overline{\langle\sigma_z\rangle^{(r)}}\).
\textbf{(b)} Trajectory variance \(\mathrm{Var}_{\mathrm{traj}}[\langle\sigma_z\rangle^{(r)}]\).
The projective unraveling is shown as a solid blue curve with diamond markers, and the kick unraveling as a dashed orange curve with triangle markers. Curves show Qiskit local simulations and markers show quantum processor data. Vertical dotted lines indicate the interventions at \(t_1\) and \(t_2\). Error bars are obtained via bootstrap resampling. Parameters: \(\Omega=4\), \(\Delta=2\), \(t_1=2\), \(t_2=4\), \(t_f=5\); 1000 shots per time point.}
 \label{fig:1Q_means_var}
\end{figure}

Figure~\ref{fig:1Q_means_var} shows the time dependence of the trajectory-averaged expectation value \(\overline{\langle \sigma_z \rangle^{(r)}}\) and the trajectory variance \(\mathrm{Var}_\mathrm{traj}[\langle \sigma_z \rangle^{(r)}]\) for both unravelings. Solid curves show Qiskit local simulations (checked against direct theoretical calculations), and markers show quantum processor data. The projective and kick unravelings coincide at the level of \(\overline{\langle \sigma_z \rangle^{(r)}}\), but they separate clearly at the level of the trajectory variance.

\begin{figure}[t]
 \centering
 \includegraphics[width=\linewidth]{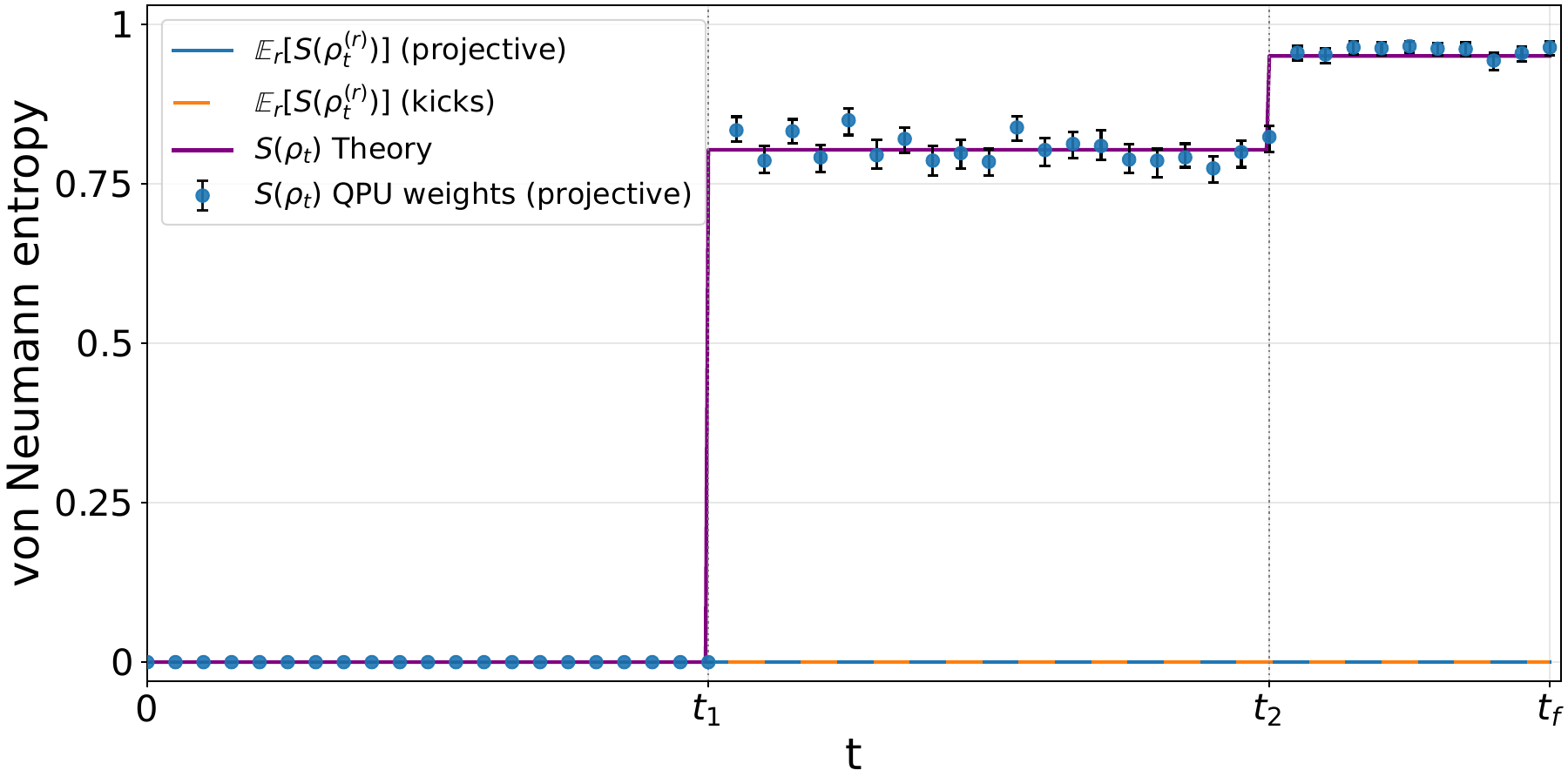}
\caption{Single-qubit protocol: entropy-based quantities as functions of time (von Neumann entropy computed with a base-2 logarithm). Purple: \(S(\rho_t)\) for the ensemble-averaged state. Blue and orange: trajectory-averaged entropy \(\mathbb{E}_r[S(\rho_t^{(r)})]\) for the projective and kick unravelings. Blue markers: estimate of \(S(\rho_t)\) obtained by combining the theoretical state reconstruction with branch weights extracted from projective-hardware data. Error bars are obtained via bootstrap resampling of the shot counts at each time point. Vertical dotted lines indicate the interventions at \(t_1\) and \(t_2\). Parameters and intervention times are the same as in Fig.~\ref{fig:1Q_means_var}.}
 \label{fig:1Q_entropy}
\end{figure}

We also examine von Neumann entropy. The entropy of the unconditional state, \(S(\rho_t)\), depends only on the average state and is therefore unraveling blind. In contrast, the trajectory-averaged entropy \(\mathbb{E}_r[S(\rho_t^{(r)})]\) is a nonlinear function of the conditional state and generally depends on the unraveling.
In the present single-qubit protocol, however, the conditional state remains pure along each trajectory for both unravelings. As a result, \(S(\rho_t^{(r)})=0\). While the trajectory-averaged entropy vanishes for both unravelings, it stands in stark contrast to the entropy of the average state $\rho_t$, which increases when $\rho_t$ becomes mixed due to dephasing. In Fig.~\ref{fig:1Q_entropy} we compare the theoretical prediction for \(S(\rho_t)\) with an estimate obtained by constructing the theoretical mixed state using branch weights extracted from the experimental counts of the projective unraveling. The two curves track each other over time, indicating that the inferred average evolution is consistent with the effective dephasing description.

Taken together, the single-qubit results play two roles. First, they show that \(\mathrm{Var}_\mathrm{traj}[\langle \sigma_z \rangle^{(r)}]\) distinguishes unravelings with the same average dynamics, and that the distinction is visible on quantum hardware. Second, they validate the experimental pipeline used below in the two-qubit case.

\subsection{Two-qubit protocol: theory versus experiment}\label{subsec:2Q_results}

We now turn to the two-qubit protocol of Sec.~\ref{subsec:2Q_synthetic}. As in the single-qubit case, the projective and kick implementations share the same average evolution but differ in their trajectory statistics. The two-qubit setting introduces a crucial new element: entanglement. Since the ideal trajectory states of the full two-qubit system remain pure, the entropy of the reduced state, \eqref{eq:ent2}, becomes a probe of entanglement at the trajectory level.

\begin{figure}[t]
 \centering
 \includegraphics[width=\linewidth]{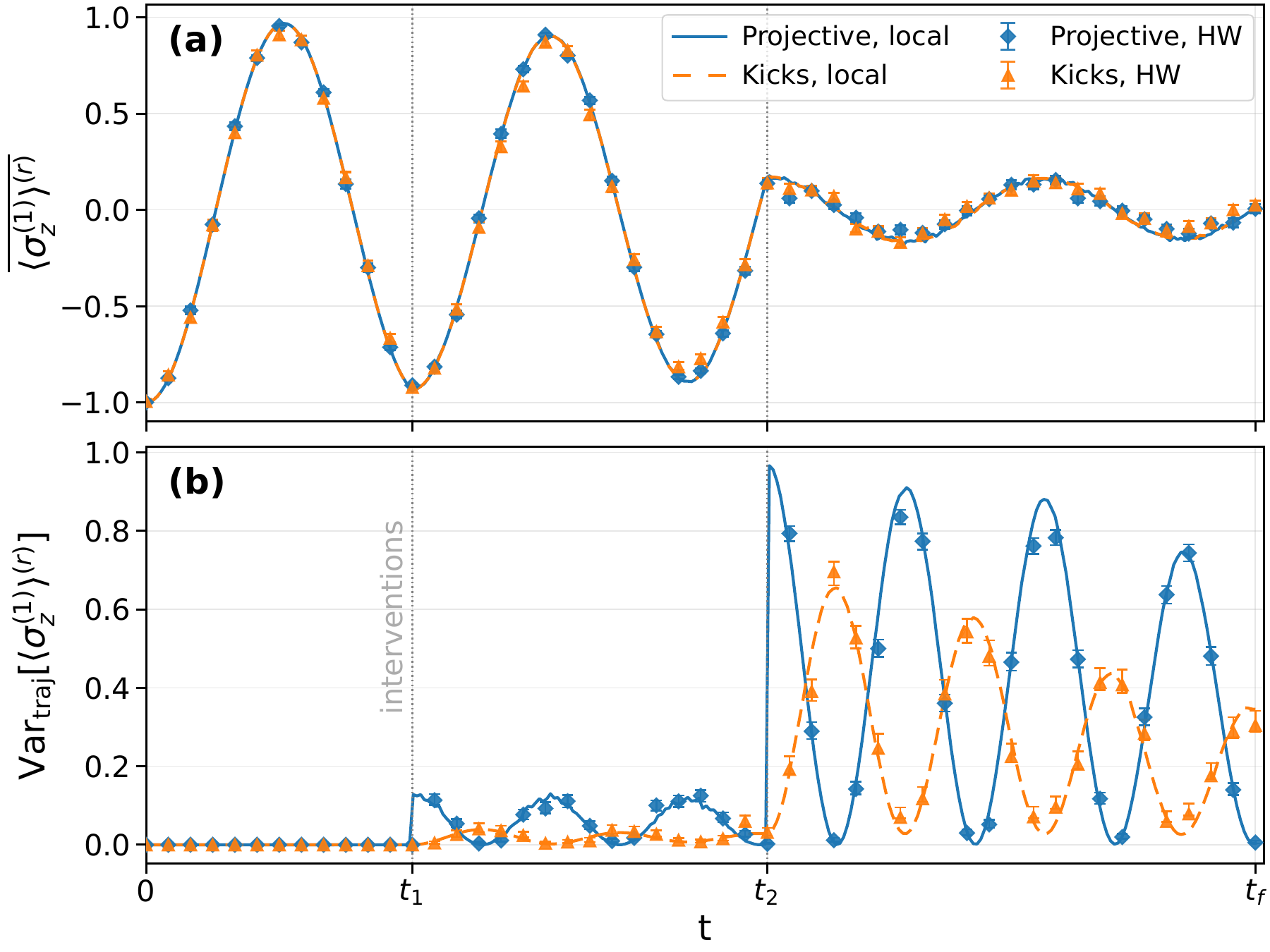}
\caption{Two-qubit protocol.
\textbf{(a)} Trajectory-averaged expectation value \(\overline{\langle\sigma_z^{(1)}\rangle^{(r)}}\).
\textbf{(b)} Trajectory variance \(\mathrm{Var}_\mathrm{traj}[\langle\sigma_z^{(1)}\rangle^{(r)}]\).
The projective unraveling is shown as a solid curve with diamond markers, and the kick unraveling as a dashed curve with triangle markers. Curves show Qiskit local simulations and markers show quantum processor data with readout-error mitigation. Vertical dotted lines indicate the interventions at \(t_1\) and \(t_2\). Error bars are obtained via bootstrap resampling. Parameters: \(\Omega = 10\), \(\Delta = 1\), \(J = -0.5\), \(t_1 = 0.6\), \(t_2 = 1.4\), \(t_f = 2.5\); 2000 shots per time point.}
 \label{fig:2Q_means_var}
\end{figure}

Figure~\ref{fig:2Q_means_var} shows \(\overline{\langle \sigma_z^{(1)} \rangle^{(r)}}\) and \(\mathrm{Var}_\mathrm{traj}[\langle \sigma_z^{(1)} \rangle^{(r)}]\) for both unravelings. The average expectation value agrees between unravelings, and the experimental data follows the simulations. 
The trajectory variances differ between the unravelings, and the separation remains clearly visible in the experimental data from quantum processor.

We next consider entropy-based quantities. Since both unravelings implement the same dephasing map at the level of the average state, the entropies \(S(\rho_t)\) and \(S(\rho_t^{(1)})\) are identical for the projective and kick implementations. At the trajectory level the global state \(\rho_t^{(r)}\) remains pure, but the reduced state
\(
 \rho_{t}^{(1,r)}=\mathrm{Tr}_2\,\rho_t^{(r)}
\)
is generally mixed, and its entropy quantifies entanglement with qubit~2 along trajectory \(r\). The trajectory-averaged reduced entropy \(\mathbb{E}_r[S(\rho_t^{(1,r)})]\) is therefore unraveling dependent.

\begin{figure}[t]
 \centering
 \includegraphics[width=\linewidth]{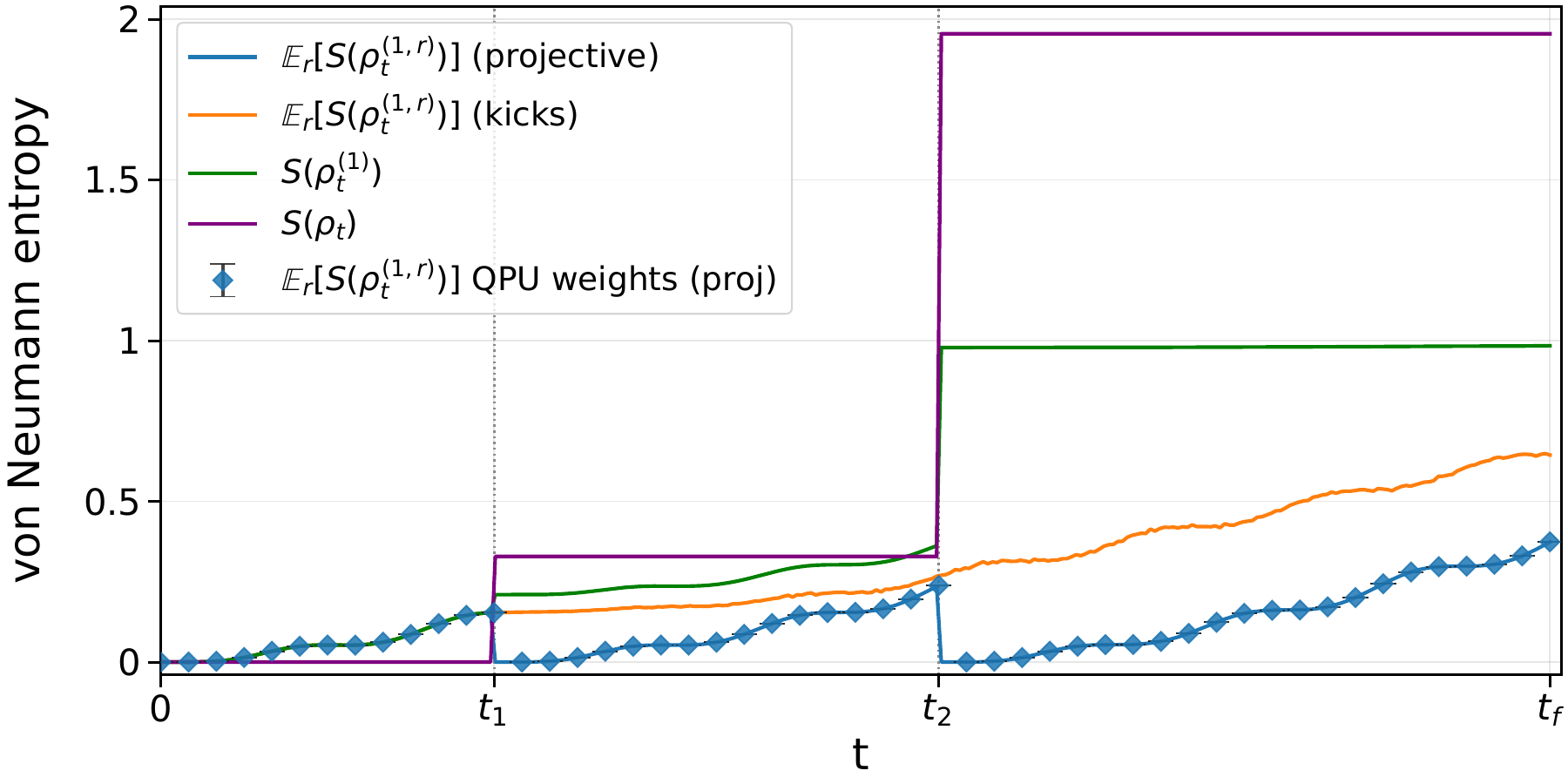}
\caption{Two-qubit protocol: entropy-based quantities (base-2 logarithm) as functions of time (same parameters and intervention times as in Fig.~\ref{fig:2Q_means_var}). Purple: \(S(\rho_t)\) for the ensemble-averaged two-qubit state. Green: \(S(\rho_t^{(1)})\) for the ensemble-averaged reduced state of qubit~1. Blue and orange: trajectory-averaged reduced entropy \(\mathbb{E}_r[S(\rho_t^{(1,r)})]\) for the projective and kick unravelings. Diamond markers: estimate obtained by combining theoretical state reconstructions with branch weights extracted from projective-hardware data (with readout-error mitigation). Error bars are obtained via bootstrap resampling of the shot counts at each time point. The trajectory-averaged reduced entropy depends on the unraveling and complements the variance-based signatures in Fig.~\ref{fig:2Q_means_var}.}
 \label{fig:2Q_entropy}
\end{figure}

Figure~\ref{fig:2Q_entropy} shows the theoretical curves for \(S(\rho_t)\), \(S(\rho_t^{(1)})\), and \(\mathbb{E}_r[S(\rho_t^{(1,r)})]\) for both unravelings. We also show points obtained by computing the entropy of the corresponding theoretical state using branch weights extracted from the projective-hardware data (with readout mitigation). The trajectory-averaged reduced entropy separates the two unravelings and is also distinct from the entropies of the average states, in line with the variance-based signatures.

Together with the findings reported in Sec.~\ref{subsec:1Q_results}, these results show the same pattern in one- and two-qubit settings: quantities that depend only on the average state are independent of the unraveling choice, while nonlinear trajectory averages remain unraveling sensitive and can be measured on current quantum processors.

\section{Discussion and outlook}\label{sec:discussion}

Our results fit into a broader line of work emphasizing that unravelings encode information not fixed by the master equation alone~\cite{Brown25gauge}. 
While the master equation strictly constrains the linear average evolution, it does not determine the stochastic dynamics at the trajectory level. Trajectory-ensemble averages of nonlinear functions probe higher moments of the distribution of conditional states and can therefore discriminate between unravelings even when all trajectory-ensemble averages of linear functions coincide.

This separation carries an important practical implication for experimental post-processing. For quantities that are linear in the conditional state, averaging the measurement record is mathematically equivalent to evaluating the same quantity on the ensemble-averaged state. For nonlinear functions, however, the order of operations matters. As our entropy results illustrate, the average of the trajectory entropies is distinct from the entropy of the average state. This distinction serves as a cautionary note for experimental characterization: estimating quantities such as the von Neumann entropy or higher-order correlation functions (e.g., $g^{(2)}(\tau)$) requires a deliberate choice of whether to probe fluctuations of individual trajectories or properties of the ensemble-averaged density matrix.
The latter does not provide access to the conditional information that characterizes a specific monitoring or intervention scheme.

A key feature of our approach is the shift from monitoring a natural environment to actively controlling the source of stochasticity. In standard quantum optics, the interaction with the electromagnetic field is intrinsic; altering the unraveling often implies a change in physical hardware (e.g., switching from photon counting to homodyne detection). In contrast, our digital protocol implements the unraveling via programmable interventions. This flexibility allows us to realize schemes like the ``kick'' unraveling—where measurement collapse is replaced by unitary randomization—that would be non-trivial to engineer in a standard experimental quantum optical platform. In fact, a more elaborate synthetic protocol could be engineered to emulate more closely a physical unraveling, such as that associated with direct photodetection in resonance fluorescence, with the key difference that fluorescence jumps reset the emitter to \(\lvert 0\rangle\), whereas the synthetic measurement may yield either \(\lvert 0\rangle\) or \(\lvert 1\rangle\).

Controlling the unraveling can also offer practical advantages for classical simulation, particularly in quantum many-body settings~\cite{VovkPichler2024, Verstraelen2023, Chen24optimized, Ruben25,Cichy25}. In particular, selecting an unraveling that suppresses entanglement of the conditional state can substantially improve the efficiency of tensor-network simulation methods~\cite{Paeckel19}. Access to trajectory-resolved observables is, moreover, essential for studying measurement-induced phase transitions (MIPTs)~\cite{Skinner2019MIPT, Li2018Zeno, Li2019Hybrid, Chan2019UnitaryProjective, Turkeshi2021InfiniteToZero, Sierant2022BeyondMultifractality, Sierant2022StabilizerDP1}.  In these transitions, the entanglement structure of individual quantum trajectories changes sharply as a function of the measurement rate, even though the corresponding unconditional mixed-state evolution may remain smooth and display no critical behavior.  The experimental ingredients used in our two-qubit demonstration---mid-circuit measurements, record-resolved data processing, and nonlinear post-processing---are precisely the building blocks required to extend our trajectory-based study to multi-qubit MIPTs~\cite{Koh2023}.

Our demonstration also highlights a central conceptual point: the \emph{same} unconditional evolution can admit inequivalent unravelings with qualitatively different trajectory physics. 
In the context of MIPTs, this implies that the presence (or absence) of a trajectory-level transition may depend on the chosen unraveling~\cite{Piccitto2022IsingUnravelings, Kolodrubetz2023Optimality, Piccitto24, Eissler25}, although certain setups have been argued to exhibit robust features across distinct unraveling classes~\cite{Niederegger25absence}. 
Furthermore, measurement-conditioned feedback provides an additional control knob: by appropriately choosing the unitary operations applied in response to the measurement outcomes, one can reshape the trajectory ensemble and, consequently, its entanglement properties~\cite{Viviescas10,Zhang17}.
With appropriately designed feedback protocols, transitions in the average state $\rho_t$ can even be brought into one-to-one correspondence with trajectory-level entanglement transitions~\cite{Sierant2023ControllingAbsorbing,Iadecola25concominant}, albeit in a way that generally requires careful engineering of the feedback mechanism~\cite{Iadecola2023ChaosControl, ODea2024InteractiveAbsorbing, Ravindranath2023Steering}.

\section{Conclusions}\label{sec:conclusions}

We have introduced and implemented single- and two-qubit protocols in which two inequivalent synthetic unravelings realize the same average dynamics. Implemented on IBM Quantum hardware, a projective scheme and a random-unitary (kick) scheme yield identical ensemble-averaged evolution and identical trajectory-ensemble averages of linear functions of the conditional state, yet display distinct nonlinear trajectory statistics.

Our experiments explicitly illustrate the operational distinction between average-state properties and trajectory averages. While for quantities that are linear in the conditional state averaging the results is equivalent to evaluating the observable on the average state, our data confirms that for nonlinear functions these two procedures generally differ. By accessing the conditional states, we showed that the specific measurement backaction—invisible to the average dynamics—can be resolved experimentally through appropriate nonlinear post-processing.

In the single-qubit case, this separation was evidenced by the trajectory variance $\mathrm{Var}_\mathrm{traj}[\langle\sigma_z\rangle^{(r)}]$, which clearly distinguished the unravelings despite the agreement of the mean $\overline{\langle\sigma_z\rangle^{(r)}}$. In the interacting two-qubit protocol, we supplemented the results for variances by entropic quantities: the trajectory-averaged reduced entropy proved to be a robust signature of the unraveling, providing complementary information to the variance.

Looking ahead, the same trajectory-resolved toolbox provides a natural route to scaling these ideas to larger systems. In particular, it establishes the experimental basis for investigating measurement-driven entanglement phenomena that rely on conditional physics, such as measurement-induced phase transitions.
    
\backmatter

\bmhead{Acknowledgements}
We acknowledge discussions with B.~Andrade on IBM Quantum devices and their use for this work.

E.P. acknowledges support from the Spanish Ministry of Science and Innovation through the predoctoral fellowship PRE2021-098926, funded by MCIN/AEI/10.13039/501100011033 and by the European Social Fund Plus (FSE+).

P.S. acknowledges fellowship within the “Generación D” initiative, Red.es, Ministerio para la Transformación Digital y de la Función Pública, for talent attraction (C005/24-ED CV1), funded by the European Union NextGenerationEU funds, through PRTR.

The ICFO Quantum Optics Theory group acknowledges support from the European Research Council (ERC) under the European Union's Horizon 2020 research and innovation programme (AdG NOQIA); from MCIN/AEI/10.13039/501100011033 through projects PGC2018-0910.13039/501100011033, CEX2019-000910-S/10.13039/501100011033, Plan Nacional FIDEUA PID2019-106901GB-I00, Plan Nacional STAMEENA PID2022-139099NB-I00, and the project funded by MCIN/AEI/10.13039/501100011033 and by the European Union NextGenerationEU/PRTR (PRTR-C17.I1), including FPI support; from QUANTERA project DYNAMITE PCI2022-132919, QuantERA II Programme co-funded by the European Union’s Horizon 2020 programme under Grant Agreement No.~101017733; from the Ministry for Digital Transformation and of Civil Service of the Spanish Government through the QUANTUM ENIA project call (Quantum Spain project), and by the European Union through the Recovery, Transformation and Resilience Plan -- NextGenerationEU within the framework of the Digital Spain 2026 Agenda; from Fundació Cellex; from Fundació Mir-Puig; and from the Generalitat de Catalunya (European Social Fund, FEDER and CERCA programmes).

M.W.M. and M.L. acknowledge support from ICREA (Institució Catalana de Recerca i Estudis Avançats).

M.A G-M acknowledges support from the Ministry for Digital Transformation and of Civil Service of the Spanish Government through the QUANTUM ENIA project
call—Quantum Spain project, and by the European Union through the Recovery, Transformation and Resilience Plan—NextGenerationEU within the framework of the Digital Spain 2026 Agenda: also from
Projects of MCIN with funding from European Union
NextGenerationEU (PRTR-C17.I1) and by Generalitat
Valenciana, with reference 20220883 (PerovsQuTe) and
COMCUANTICA/007 (QuanTwin).

We acknowledge support from Barcelona Supercomputing Center (MareNostrum) under grant FI-2023-3-0024.

Funded by the European Union. Views and opinions expressed are however those of the author(s) only and do not necessarily reflect those of the European Union, European Commission, European Climate, Infrastructure and Environment Executive Agency (CINEA), or any other granting authority. Neither the European Union nor any granting authority can be held responsible for them (HORIZON-CL4-2022-QUANTUM-02-SGA PASQuanS2.1, 101113690; EU Horizon 2020 FET-OPEN OPTOlogic, Grant No.~899794; QU-ATTO, 101168628; EU Horizon Europe research and innovation programme NeQST, Grant Agreement No.~101080086).

We also acknowledge support from the ICFO internal ``QuantumGaudi'' project.

We acknowledge the use of IBM Quantum services for this work. The views expressed are those of the authors and do not necessarily reflect the views of IBM or IBM Quantum.

\section*{Declarations}

\noindent\textbf{Funding} \\
This work was supported by the funding agencies and projects listed in the Acknowledgements section, including MCIN/AEI/10.13039/501100011033 and the European Union (FSE+, Horizon 2020, Horizon Europe, and NextGenerationEU/PRTR), as well as national and regional programmes and institutional support (Fundació Cellex, Fundació Mir-Puig, Generalitat de Catalunya, ICREA, Barcelona Supercomputing Center, and ICFO internal projects).

\medskip
\noindent\textbf{Competing interests} \\
The authors have no competing interests to declare that are relevant to the content of this article.

\medskip
\noindent\textbf{Ethics approval} \\
Not applicable.

\medskip
\noindent\textbf{Consent to participate / publish} \\
Not applicable.

\medskip
\noindent\textbf{Data and code availability} \\
The numerical simulation codes and experimental data generated during the current study are available from the corresponding author on reasonable request.

\medskip
\noindent\textbf{Authors' contributions} \\
The original idea, general conceptual framework and theoretical derivations were developed by P.~Sierant, E.~Pi\~nol and M.~Lewenstein. E.~Pi\~nol performed the numerical simulations, designed and implemented the quantum circuits, and carried out the IBM Quantum experiments and data analysis. All authors contributed to the interpretation of the results, to the refinement of the theoretical framework, and to the writing and revision of the manuscript and approved the final version.


\bibliography{references}

\end{document}